\documentclass[twocolumn,showpacs,preprintnumbers,prl,nofootinbib,aps,10pt]{revtex4-1}
\usepackage{graphicx,epsfig,amsmath,amssymb,bm}

\usepackage{color}
\usepackage[normalem]{ulem}

\makeatletter
\DeclareRobustCommand*{\bfseries}{%
  \not@math@alphabet\bfseries\mathbf
  \fontseries\bfdefault\selectfont
  \boldmath
}
\makeatother

\usepackage{hyperref}
\usepackage{graphics}
\usepackage{epsf}
\usepackage{subfigure}
\usepackage{mathtools}

\usepackage{multirow}
\usepackage{nicefrac}
\usepackage{epstopdf}
\usepackage{slashed}
\usepackage{xcolor}
\usepackage{feynmp}

\allowdisplaybreaks

\def\mev{{\rm MeV}}

\begin{document}

\preprint{IU-HET-597} 

\title{Emerging Lattice approach to the K-Unitarity Triangle}

\author{Christoph Lehner$^a$, Enrico Lunghi$^b$ and Amarjit Soni$^a$} 

\affiliation{${}^a$\,Physics Department, Brookhaven National Laboratory, Upton, NY 11973, USA\\
${}^b$\,Physics Department, Indiana University, Bloomington, IN 47405, USA
}

\begin{abstract}
It has  been clear for the past several years that new physics in the quark sector can only appear, in low energy observables, as a perturbation. Therefore precise theoretical predictions and precise experimental measurements have become mandatory. Here we draw attention to the significant advances that have been made in lattice QCD simulations in recent years in $K\to \pi \pi$, in the long-distance  contribution to indirect CP violation in the Kaon system ($\varepsilon$) and in rare $K$-decays.  Thus, in conjunction with experiments, the construction of a unitarity triangle purely from Kaon physics should soon become feasible. We want to emphasize that in our approach to the $K$-unitarity triangle, the ability of lattice QCD methods to systematically improve the calculation of the direct CP-violation parameter ($\varepsilon^{\prime}$) plays a pivotal role. Along with the B-unitarity triangle, this should could allow, depending on the pattern of new physics, for more stringent tests of the Standard Model and tighter constraints on new physics.
\end{abstract}

\maketitle

\section{Introduction and Motivation}
\label{sec:introduction}
For  the past decade or more, from a variety of low energy precision experiments, such as  those from B-factories and LHCb, it is becoming clear that effects of new physics are likely to just show up as perturbations. The current tests on the unitarity triangle show very good agreement with the Standard Model (SM) CKM-paradigm~\cite{Cabibbo:1963yz, Kobayashi:1973fv} to the level of about~20\%. Impressive as this is, it is also important to emphasize that 10-15\% effects of new physics (which at present can not be ruled out) are quite big. In fact it is  useful to recall that CP violation was first discovered in the decays $K_L \to \pi \pi$ at the level of $10^{-3}$~\cite{Christenson:1964fg};  therefore we should not be surprised if similar precise measurements become necessary to discover new phenomena.  Thus the need for improved precision in experiments as well as in theory should be clear.

In this work we highlight recent advances made in the lattice computation of $K$-decays that had been serious challenges for a very long time. As will be explained, in the course of tackling the calculation of the crucially important direct CP violation parameter in $K \to \pi \pi$  decays, $\varepsilon^{\prime}/\varepsilon$, by using the Lellouch-L\"uscher method involving finite-volume correlation functions, the RBC and UKQCD collaborations managed to develop an interesting method to tackle matrix elements of non-local 4-quark operators which are relevant to quantify ``Long-Distance" (LD) contributions to a variety of matrix elements of physical interest. Explicit examples under current active study are the neutral Kaon mass difference, the related long-distance part of the  indirect CP-violation parameter $\varepsilon$ and the branching ratio (BR) of the rare decay $K^+ \to \pi^+ \nu \bar\nu$. 

For $\varepsilon^{\prime}/\varepsilon$, $\varepsilon$, and $\Delta m_K$ rather precise experimental measurements already exist~\cite{Agashe:2014kda}. The initial measurement of BR$(K^+ \to \pi^+ \nu \bar\nu) =  (1.73^{+1.15}_{-1.05}) \times 10^{-10}$ was accomplished some years ago at BNL in experiment E-787 and E-949~\cite{Adler:1997am, Artamonov:2008qb}. At CERN the NA62 experiment is expected to significantly improve the determination of this branching ratio~\cite{fortheNA62:2013jsa, Rinella:2014wfa, Romano:2014xda} in the next few years.

An important point to consider is also that lattice methods for tackling non-perturbative effects are largely systematically improvable. This means once the physical quantity becomes amenable to lattice methods,  accuracy with computer capability and time is essentially guaranteed.  A few examples are the Kaon $B$-parameter $\hat B_K$ (that is the ratio of the complete $K$-$\bar K$ matrix element of the leading SM operator to its estimate in the naive factorization approximation) which in full QCD with chiral fermions got evaluated with about 7-8\% total error around 2007. In the next $\sim$ 5 years, many different collaborations have attacked it with different discretizations and the world average now has an error around 1\%.  Quark masses, $K\to \pi\ell\nu$, $K\to \pi \pi$ in the $I=2$ channel, $f_B$, $B$-mixings, and others have followed a similar path (see, {\it e.g.}, Ref.~\cite{Aoki:2013ldr} for a review of the present status of flavor physics lattice calculations). Calculations of $\varepsilon^{\prime}/\varepsilon$ and the long-distance effects in $K$-$\bar K$ mixings and in rare $K$ decays are expected to progress in analogous ways.
 
Bearing in mind these exciting developments in theory and in experiments, it seems timely to ask if we can now construct a unitarity triangle based primarily from input from $K$-physics, which has been often talked about~\cite{Buras:1994ec, Buchalla:1994tr, Buchalla:1996fp, Buras:2001af, Buras:2006gb, Blucher:2009zz},  and is the subject of this work.  

By way of motivation, let us recall that, in general, naturalness arguments suggest that beyond the SM (BSM) scenarios are unlikely to be flavor blind. Indeed, just as in the SM, as weak interactions are ``switched-on" the gauge eigenstates are no longer aligned with the mass eigenstates and the connection between the two is monitored by the CKM-matrix. Warped extra dimensions provide a very interesting example, as they provide a geometric  understanding of flavors. Many aspects of flavor physics can be readily understood through localization of different flavors at different locations in the extra-dimension yielding a non-universal,  strongly mass-dependent effect on flavors accompanied by many O(1) BSM-phases~\cite{Agashe:2004cp, Blanke:2008zb, Casagrande:2010si}. This is a very illustrative example, and in many if not most BSMs a similar situation arises.

Another important consideration by way of motivation is that the observables in the $K$-unitarty triangle are widely believed to be very sensitive to deviations from minimum flavor violation (MFV); this is in contrast to the many observables used in the standard $B$-unitarity triangle. Thus using the $B$-UT to extract SM-CKM parameters and then the 	$K$-observables for new physics searches can be a very effective approach.

The ability of these new lattice methods to quantify non-perturbative effects in $K$-$\pi$ physics calls for a re-examination of Kaon experiments. For one thing given that in a few years the theory errors on $\varepsilon^{\prime}$ are likely to come down to around 11\% of the current experimental central value with the real possibility of further improvements down the road, improved experimental determinations may be called for given the current experimental error is around 15\%.

Moreover, although measurement of $\varepsilon^{\prime}$ is very challenging, measuring the rate for $K_L \to \pi^0 \nu \bar \nu$ is even more challenging. Traditionally, this very difficult purely CP-violating rare mode~\cite{Littenberg:1989ix} has been a pristine SM prediction as it is clearly short-distance dominated because of the large top quark mass. However, as lattice methods make sufficient progress in $\varepsilon^{\prime}$, the theoretical advantage of $K_L \to \pi^0 \nu \bar \nu$ will likely diminish. Note, however, that contributions to these two processes are independent in many models making both invaluable tools in searches for new physics effects. Moreover, it must be stressed that progress towards a precise determination of ${\rm BR} (K^+\to \pi^+\nu\bar\nu)$ is guaranteed by existing and planned experiments and theoretical progress in lattice QCD calculations of such processes (This will be achieved by reduced errors on $|V_{cb}|$ from improved determinations of the $B\to D$ form factor (See for instance Ref.~\cite{Lattice:2015rga} where the $q^2$ dependence of this form factor is calculated) and by a direct calculation of long distance charm contributions to ${\rm BR} (K^+\to \pi^+\nu\bar\nu)$~\cite{Feng:2015kfa}). Needless to say, improvements on the measurement of $\varepsilon^\prime/\varepsilon$ will require a major experimental effort.

Another very  interesting decay mode that, in principle, can provide a useful test of CP-violation is  $K_L \to \pi^0 e^+ e^-$ (see, {\it e.g.}, Ref.~\cite{Buras:1994qa}) which in fact allows for a detailed study via a Dalitz plot. The major theoretical distinction with $K_L \to \pi^0 \nu \bar \nu$ is that the charged lepton mode receives a non-negligible contribution to the BR from CP-conserving 2-photon intermediate state, as opposed  to $K_L \to \pi^0 \nu \bar \nu$ which, to an excellent approximation in the SM is purely CP-violating.

If theory could reliably and precisely predict the CP-conserving contribution to $K_L \to \pi^0 e^+ e^-$ which is dominated by LD effects, this mode could become extremely significant as from an experimental perspective the  $e^+ e^-$ final state is seemingly a lot easier to address. Unfortunately there is a daunting experimental background, pointed out in~\cite{Greenlee:1990qy} from $K_L \to \gamma \gamma e^+ e^-$ which will need to be overcome~\cite{LL}.

The measurements that we include in the Kaon unitarity triangle fit (KUT) are $\varepsilon$, $\varepsilon^\prime/\varepsilon$, ${\rm BR} (K^+\to\pi^+\nu\bar\nu)$ and $|V_{cb}^{}|$ (from inclusive and exclusive semileptonic $b\to c\ell\nu$ decays). The standard unitarity triangle fit (SUT) depends on $|V_{cb}|$ only via the ratio $|V_{ub}/V_{cb}|$ and, indirectly, via the rare decay $B\to \tau \nu_\tau$. Presently $|V_{cb}|$ provides a subdominant contribution to the uncertainties on these two quantities (the former is controlled by theoretical and experimental errors on $b\to u \ell\nu$ decays, the latter is dominated by experimental errors). Hence the SUT and KUT fits are essentially independent and any tension between them is an indication of BSM physics. Finally we note that $\Delta m_K$, while being very sensitive to the chirality of new physics~\cite{Beall:1981ze}, does not provide any constraint on the $\bar\rho -\bar\eta$ plane and is therefore not included in the KUT fit.

The structure of this manuscript is as follows: We first briefly review new lattice methodology that enabled some of the recent advances in lattice Kaon physics.  We then specifically summarize $K\to\pi\nu\bar\nu$ decays, $\varepsilon^\prime/\varepsilon$, and $\varepsilon$.  We finally present our findings for current and potential future Kaon unitarity triangle fits.

\section{Lattice Methodology}
\label{sec:methodology}
Relations between finite-volume Euclidean correlation functions, which are accessible to lattice QCD, and infinite-volume matrix elements form the foundation of many of the lattice efforts mentioned in this work.  The simplest case of one particle decaying into a single two-particle final-state allows for the extraction of $K\to \pi\pi$ matrix elements \cite{Lellouch:2000pv,Kim:2005gf}.  Extensions for matrix elements beyond this limit have recently received a lot of attention, see Ref.~\cite{Briceno:2014pka} for a summary.  For the current work, the development of methodology for non-local (bi-local) matrix elements is of particular interest \cite{Christ:2010gi,Bai:2014cva,Christ:2015pwa}. The $K_L$ -- $K_S$ mass difference computation can serve as a nice introduction to the general methodology \cite{Christ:2010gi}.  The desired infinite-volume quantity
\begin{align}
  \Delta m_K &= 2{\cal P} \sum_n \frac{ \langle \overline{K}^0 \vert H_W \vert n \rangle \langle n \vert H_W \vert K^0 \rangle }{ m_K - E_n }
\end{align}
can be related to its finite-volume counter-part
\begin{align}
  \Delta m_K^{\rm FV} &= 2 \sum_{n \neq n_0} \frac{ \langle \overline{K}^0 \vert H_W \vert n \rangle \langle n \vert H_W \vert K^0 \rangle }{ m_K - E_n }
\end{align}
with controlled finite-volume errors, where it is assumed that a single $\pi$--$\pi$ intermediate state $n_0$ is degenerate with $K^0$ and $\overline{K}^0$.   The remaining task is to extract $\Delta m_K^{\rm FV}$ from Euclidean space correlation functions which is usually formulated in terms of four-point functions
\begin{align}
  {\cal A} &= \frac12 \langle \overline{K}^0(t_f) \int_{t_a}^{t_b} dt_2\int_{t_a}^{t_b} dt_1 H_W(t_2) H_W(t_1) K^0(t_i) \rangle \,.
\end{align}
This amounts to the summation of operator insertions within a fiducial volume bounded by time coordinates $t_a$ and $t_b$ with source and sink operators at $t_i$ and $t_f$ satisfying $t_f \ll t_a < t_b \ll t_i$. Inserting a full set of states and performing the integrals yields
\begin{align}
  {\cal A} = & - \sum_{n \neq n_0} \frac{ \langle \overline{K}^0 \vert H_W \vert n \rangle \langle n \vert H_W \vert K^0 \rangle }{ m_K - E_n }
  \Bigg[ t_b - t_a  \nonumber \\
&  - \frac{e^{-(E_n-m_K)(t_b-t_a)} - 1}{m_K - E_n}\Bigg] e^{-(t_f-t_i)m_K} \\
&  - \frac12 (t_b - t_a)^2  \langle \overline{K}^0 \vert H_W \vert n_0 \rangle \langle n_0 \vert H_W \vert K^0 \rangle e^{-(t_f - t_i)m_K} \,. \nonumber
\end{align}
The coefficient of $t_b-t_a$ is the desired $\Delta m_K^{\rm FV}$.  For states $n$ with $E_n < m_K$ there are exponentially growing contributions that complicate the extraction of the linear $t_b-t_a$ dependence. The control of these exponentially growing terms makes it clear that the methodology works best if the number of such contributions is small. A variant of this general procedure can be used to compute long-distance contributions to $\varepsilon$ and rare Kaon decays mentioned in this work.  Work along those lines is in progress~\cite{Christ:2014qwa,Feng:2015kfa}.

\section{The rare decay \texorpdfstring{$K\to \pi\nu\bar\nu$}{}}
\label{sec:kpnn}
The branching ratios for the $K^+\to\pi^+\nu\bar\nu$ and $K_L \to \pi^0 \nu\bar \nu$ decays are given by~\cite{Haisch:2007pd, Buras:2015qea, Knegjens:2015gea}:
\begin{align}
{\rm BR} (K^+\to\pi^+\nu\bar\nu (\gamma)) = \; & 
\kappa_+ (1+ \Delta_{\rm EM}) \Bigg[ \nonumber \\
& \hskip -2.6cm  \left( \frac{{\rm Im} \left( V_{td}^{} V_{ts}^* \right)}{\lambda^5}\; X(x_t) \right)^2 \nonumber \\
& \hskip -3cm + \left( \frac{{\rm Re} \left( V_{cd}^{} V_{cs}^* \right)}{\lambda}\; P_c(X) + 
       \frac{{\rm Re} \left( V_{td}^{} V_{ts}^* \right)}{\lambda^5}\; X(x_t)
\right)^2 \Bigg] \; , \\
{\rm BR} (K_L\to\pi^0\nu\bar\nu) = \; & 
\kappa_L 
\left( \frac{{\rm Im} \left( V_{td}^{} V_{ts}^* \right)}{\lambda^5}\; X(x_t) \right)^2 \label{eq:kpnn}
\end{align}
where
\begin{align}
\kappa_+ &= (5.173 \pm 0.025) \cdot 10^{-11} \left[\frac{\lambda}{0.2252}\right]^8~\text{\cite{Mescia:2007kn}}\; ,\\
\kappa_L &= (2.231 \pm 0.013) \cdot 10^{-11} \left[\frac{\lambda}{0.2252}\right]^8~\text{\cite{Mescia:2007kn}}\; ,\\
\Delta_{\rm EM} &= -0.003~\text{\cite{Mescia:2007kn}}\; ,\\
X(x_t) &= 1.481 \pm 0.005_{\rm th} \pm 0.008_{\rm exp}~\text{
\cite{Buchalla:1993bv, Misiak:1999yg, Brod:2010hi, Buras:2015qea}}\; ,\\
P_c(X) &= (0.404 \pm 0.024)\left[\frac{\lambda}{0.2252}\right]^4 ~\text{
\cite{Buchalla:1998ba, Buchalla:1993wq, Buras:2005gr, Buras:2006gb, Isidori:2005xm, Brod:2008ss}}\; .
\end{align}
The theoretical results for these two decays read (keeping an explicit dependence on the CKM angles)~\cite{Buras:2015qea}:
%
\begin{align}
{\rm BR} (K^+\to\pi^+ \nu\bar\nu) & \; = \Big\{
\left(8.39 \pm 0.30 \right) \; \times 10^{-11}  \nonumber\\
& \hskip -1cm \times \left[ \frac{|V_{cb}|}{40.7 \times 10^{-3}} \right]^{2.8}
\left[ \frac{\gamma}{73.2^{\rm o}} \right]^{0.708}\Big\} \; ,  \label{eq:kpnnBURAS}\\
{\rm BR} (K_L\to\pi^0 \nu\bar\nu) & \; 
= \Big\{\left(3.36 \pm 0.05 \right) \; \times 10^{-11}  \nonumber\\ 
&  \hskip -3cm \times
\left[ \frac{|V_{ub}|}{3.88 \times 10^{-3}} \right]^2
\left[ \frac{|V_{cb}|}{40.7 \times 10^{-3}} \right]^2 \left[ \frac{\sin(\gamma)}{\sin(73.2^{\rm o})} \right]^2\Big\} .
\end{align}
%
Note that the explicit CKM dependence in Eq.~(\ref{eq:kpnnBURAS}) is accurate to about 1\% in the ranges $37 \times 10^{-3} < |V_{cb}| < 45 \times 10^{-3}$ and $60^{\rm o} < \gamma < 80^{\rm o}$. The non-CKM uncertainty on ${\rm BR} (K_L \to \pi^0\nu\bar\nu)$ is much smaller than in the charged mode due to large uncertainties associated with the quantity $P_c(X)$ introduced in Eq.~(\ref{eq:kpnn}). 

Using the complete unitarity triangle fit results to determine the relevant CKM entries, we obtain the following SM predictions:
\begin{align}
{\rm BR} (K^+\to\pi^+ \nu\bar\nu) & \; =
\begin{cases}
(8.64 \pm 0.60) \times 10^{-11}  & \hskip -0.2cm \text{SM} \cr
\left(17.3^{+11.5}_{-10.5}\right) \times 10^{-11} & \hskip -0.2cm \text{E949~\cite{Artamonov:2008qb}} \cr
\end{cases} \nonumber \\
{\rm BR} (K_L\to\pi^0 \nu\bar\nu) & \; =
\begin{cases}
(2.88 \pm 0.25 ) \times 10^{-11} &  \hskip -0.2cm \text{SM} \cr
< 2.6 \times 10^{-8} &  \hskip -0.2cm \text{E391a~\cite{Ahn:2009gb}}\cr
\end{cases} \nonumber
\end{align}
In the study presented below we consider two future scenarios in which the experimental central value of the $K^+\to\pi^+ \nu\bar\nu$ branching ratio either does not change or shifts to the SM prediction. The NA62 experiment should be able to reduce the uncertainties to $\delta_{\rm exp} = 7\%$ and $10\%$, respectively. (These estimates are based on the expectation that NA62 will collect about 100 SM events by 2017~\cite{Rinella:2014wfa, Romano:2014xda}; the 7\% uncertainty is obtained by rescaling the expected number of events by the ratio ${\rm BR} (K^+\to\pi^+\nu\bar\nu)_{\rm exp}/{\rm BR} (K^+\to\pi^+\nu\bar\nu)_{\rm SM}$.) The KOTO experiment at JPARC expects to observe $K_L\to \pi^0 \nu\bar \nu$ at the SM level with a 10\% uncertainty~\cite{Yamanaka:2012yma, Komatsubara:2012pn, Shiomi:2014sfa}. 

\section{Direct CP-violation in \texorpdfstring{$K\to\pi\pi$}{} and \texorpdfstring{$\varepsilon^\prime/\varepsilon$}{}}
\label{sec:epoe}
The effective Hamiltonian responsible for contributions to $\varepsilon^\prime/\varepsilon$ is, at scales larger than $\mu \sim O( m_c )$~\cite{Buras:1993dy}:
%
\begin{align}
{\cal H}_{\rm eff} = \frac{G_F}{\sqrt{2}} \left[ 
\lambda_t \; \sum_{i=1}^{10} C_i  Q_i   
+ \lambda_c  \;
\left( \sum_{i=1}^2 C_i  ( Q_i  - Q_i^c  ) \right)
\right]\;,
\label{heff}
\end{align}
where $\lambda_q = V^*_{qs} V^{}_{qd}$ and the current--current, penguin and semi--leptonic operators are:
%
\begin{align}
Q_1^c   & =  (\bar{s}_\alpha c_\beta)_{V-A} (\bar{c}_\beta d_\alpha)_{V-A} \,, \\
Q_2^c   & =  (\bar{s} c)_{V-A} (\bar{c} d)_{V-A}  \,,\\
Q_1   & =  (\bar{s}_\alpha u_\beta)_{V-A} (\bar{u}_\beta d_\alpha)_{V-A}  \,,\\
Q_2   & =  (\bar{s} u)_{V-A} (\bar{u} d)_{V-A} \,, \label{eq:q1q2} \\
Q_3   & =  (\bar{s} d)_{V-A} \sum (\bar{q} q)_{V-A}  \,,\\
Q_4   & =  (\bar{s}_\alpha d_\beta)_{V-A} \sum (\bar{q}_\beta      q_\alpha)_{V-A}  \,,\\
Q_5   & =  (\bar{s} d)_{V-A} \sum (\bar{q}      q)_{V+A}  \,,\\
Q_6   & =  (\bar{s}_\alpha d_\beta)_{V-A} \sum (\bar{q}_\beta      q_\alpha)_{V+A}  \,,\\
Q_{7} & =  \frac{3}{2} (\bar{s} d)_{V-A} \sum e_q (\bar{q} q)_{V+A}  \,,\\
Q_{8} & =  \frac{3}{2} (\bar{s}_\alpha d_\beta)_{V-A} \sum e_q (\bar{q}_\beta q_\alpha)_{V+A} \,, \\
Q_{9} & =  \frac{3}{2} (\bar{s} d)_{V-A} \sum e_q (\bar{q} q)_{V-A} \,, \\
Q_{10} & =  \frac{3}{2} (\bar{s}_\alpha d_\beta)_{V-A} \sum e_q (\bar{q}_\beta q_\alpha)_{V-A} \,, \label{eq:q9q10}
\end{align}
%
where the sums in the operators $Q_{3-10}$ run over the four lightest quarks, $e_q$ denotes the quark electric charges, $\alpha$ and $\beta$ denote color indices (omitted for color singlet operators), and $(V\pm A) = \gamma_\mu (1\pm \gamma_5)$. Below $\mu \sim O(m_c)$ the charm quark is integrated out and the effective Hamiltonian is more commonly written as:
\begin{align}
{\cal H}_{\rm eff} &= \frac{G_F}{\sqrt{2}} \lambda_{u} \sum_{i=1}^{10}
\left[ z_i (\mu) + \tau \; y_i (\mu) \right] \; Q_i (\mu)\,,
\label{heff-belowmc}
\end{align}
where the operators $Q_{1,2}^c$ have disappeared, the sums in the operators $Q_{3-10}$ run over the three lightest quarks, and $\tau = -\lambda_{t}/\lambda_{u} = -(V^*_{ts} V^{}_{td})/(V^*_{us} V^{}_{ud})$ (note that $\lambda_u + \lambda_c + \lambda_t= 0$). Complete NLO expressions for the coefficients $z_i (\mu)$ and $y_i (\mu)$ can be found in Refs.~\cite{Buras:1993dy, Buchalla:1995vs}. 

\begin{table}
\begin{center}
\begin{tabular}{|c|c|c|} \hline
$i$ & $\langle Q_i \rangle_0 $ & $\langle Q_i \rangle_2$  \\
\hline
1 & -0.151(44) & 0.00965(59) \\
2 & 0.169(56) & 0.00965(59) \\
3 & -0.0492(661) & 0 \\ 
4 & 0.271(111) & 0 \\
5 & -0.191(64) & 0 \\
6 & -0.379(128) & 0 \\
7 & 0.219(61) & 0.286(11) \\
8 & 1.72(39) & 1.314(76) \\
9 & -0.202(70) & 0.01447(89) \\
10 & 0.118(50) & 0.01447(89) \\ \hline
\end{tabular}
\caption{Current determinations of $K\to (\pi\pi)_{I=0,2}$ matrix elements for each of the operators given in Eqs.~(\ref{eq:q1q2}--\ref{eq:q9q10}). For $I=0$ and $I=2$ we take $\mu=1.531 \; {\rm GeV}$ and $\mu = 3 \; {\rm GeV}$, respectively. \label{tab:bi}}
\end{center}
\end{table}
Following Refs.~\cite{Buras:1993dy, Ciuchini:1995cd, Buras:1998ra, Bosch:1999wr, Buras:2003zz}, we write:
\begin{align}
{\rm Re} \left(  \frac{\varepsilon^\prime}{\varepsilon}  \right)   = & \;
{\rm Re} \left( \frac{ \omega \; e^{i (\delta_2 - \delta_0 + \pi/2)}}{\sqrt{2}\; \varepsilon } \left[ 
\frac{{\rm Im} (A_2)}{{\rm Re} (A_2)} - \frac{{\rm Im} (A_0)}{{\rm Re} (A_0)} \right] \right) \\
=&\;  \frac{\omega}{\sqrt{2} \; |\varepsilon|} \cos \left( \delta_2 - \delta_0 + \pi/2 -\phi_\varepsilon \right) \nonumber \\
& \times \left[ \frac{{\rm Im} (A_2)}{{\rm Re} (A_2)} - \frac{{\rm Im} (A_0)}{{\rm Re} (A_0)} \right] \,,\\
A_0 =&\;  \frac{G_F}{\sqrt{2}} \lambda_{u} \sum_{i=1}^{10}
\left[ z_i (\mu) + \tau \; y_i (\mu) \right] \langle (\pi\pi)_{I=0} | Q_i | K \rangle \,, \\
A_2 = &\;  \frac{G_F}{\sqrt{2}} \lambda_{u} \sum_{i=1}^{10}
\left[ z_i (\mu) + \tau \; y_i (\mu) \right] \langle (\pi\pi)_{I=2} | Q_i | K \rangle  \,,
\end{align}
where the quantities $\omega = {\rm Re} A_2/{\rm Re}A_0 =0.04454(12)$, ${\rm Re} A_0 = 3.3201(18) \times 10^{-7}\; {\rm GeV}$, ${\rm Re} A_2 = 1.4788(41) \times 10^{-8}\; {\rm GeV}$ and 
\begin{align}\label{eqn:phieps}
  \phi_\varepsilon = (43.5 \pm 0.5)^{\rm o}  
\end{align}
are taken from experiments~\cite{Agashe:2014kda, Liu:2012}. The direct and experimental determination of the phases $\delta_{0,2}$ differ at the two sigma level
\begin{align}
\phi_{\varepsilon^\prime} = \delta_2 - \delta_0 + \frac{\pi}{2} = 
\begin{cases}
(42.3 \pm 1.5 )^{\rm o} & \text{PDG~\cite{Agashe:2014kda}} \cr
(54.6 \pm 5.8 )^{\rm o} & \text{RBC~\cite{Blum:2015ywa, Bai:2015nea}} \cr
\end{cases} \; .
\end{align}
Fortunately, due to the central value of the combination $\delta_2 - \delta_0 + \pi/2 -\phi_\varepsilon$ and to the large uncertainties in the determination of the various matrix elements, these two choices yield almost identical results; for definiteness, we follow the approach of Ref.~\cite{Bai:2015nea} and use the phases extracted from the lattice. 
\begin{table}[t]
\begin{center}
\begin{tabular}{ll}
\hline\hline
$\eta_1 = 1.87 \pm 0.76$~\cite{Brod:2011ty} & 
$m_{t, pole} = (173.5 \pm 1.0) \; {\rm GeV}$ \vphantom{\Big(}\\
$\eta_2 = 0.5765 \pm 0.0065$~\cite{Buras:1990fn}  & 
$m_c(m_c) = (1.275 \pm 0.025 ) \; {\rm GeV}$\vphantom{\Big(}\\
$\eta_3 = 0.496 \pm 0.047$~\cite{Brod:2010mj}  &  
$\varepsilon_{\rm exp} = (2.232 \pm 0.007 ) \times 10^{-3}$  \vphantom{\Big(} \vphantom{\Big(}\\
$\hat B_K = 0.766 \pm 0.010 $~\cite{Aoki:2013ldr} & 
$f_K = (156.3 \pm 0.9) \; \mev$~\cite{Aoki:2013ldr} \vphantom{\Big(} \\
\hline
\hline
\end{tabular}
\caption{Inputs used in the calculation of $\varepsilon$. The top and charm quark masses as well as quantities not explicitly given are taken from Refs.~\cite{Agashe:2014kda, Aoki:2013ldr}.\label{tab:ekinputs}}
\end{center}
\end{table}
The inclusion of isospin breaking corrections 
modifies the QCD penguins ($Q_{3-6}$) contribution to the $(\pi\pi)_{I=0}$ amplitude~\cite{Buras:2015yba}:
\begin{align}
\left[ {\rm Im} (A_0) \right]_{\rm QCDP} \to & \; \left[{\rm Im} (A_0)\right]_{\rm QCDP} \left(1-\hat \Omega_{\rm eff}\right) a \; , \label{iso1} \\
a = & \; 1.017~\text{\cite{Cirigliano:2003gt}} \; , \label{iso2} \\
\hat \Omega_{\rm eff} = & \; (14.8 \pm 8.0) \times 10^{-2}~\text{\cite{Buras:1987wc, Cirigliano:2003nn, Cirigliano:2003gt, Buras:2015yba}} \; . \label{ib}
\end{align}
As we infer from Eq.~(\ref{ib}), phenomenological estimates of isospin breaking effects vary over a considerable range and per Refs.~\cite{Buras:1987wc, Cirigliano:2003gt, Cirigliano:2003nn, Buras:2015yba} may be as high as about 15\%. In any case these corrections are likely subdominant to the  uncertainties in the current lattice calculations~\cite{Blum:2015ywa, Bai:2015nea} and are therefore being ignored at present. However, electromagnetic and isospin effects will be calculated in lattice QCD simulations in the next few years (See, for instance, Ref.~\cite{Carrasco:2015xwa} for a discussion of technical issues involved in these calculations). 

The imaginary part of the $I=0,2$ matrix elements have been recently calculated and read~\cite{Blum:2015ywa, Bai:2015nea}:
\begin{align}
{\rm Im} (A_2) & \; = \left( -6.99 \pm 0.20_{\rm stat} \pm 0.84_{\rm syst} \right) \times 10^{-13} \; {\rm GeV} \; , \label{rbca2} \\
{\rm Im} (A_0) & \; = \left( -1.90 \pm 1.23_{\rm stat} \pm  1.08_{\rm syst} \right) \times 10^{-11} \; {\rm GeV} \; .\label{rbca0}
\end{align}
\begin{figure}[t]
\begin{center}
\includegraphics[width=0.99 \linewidth]{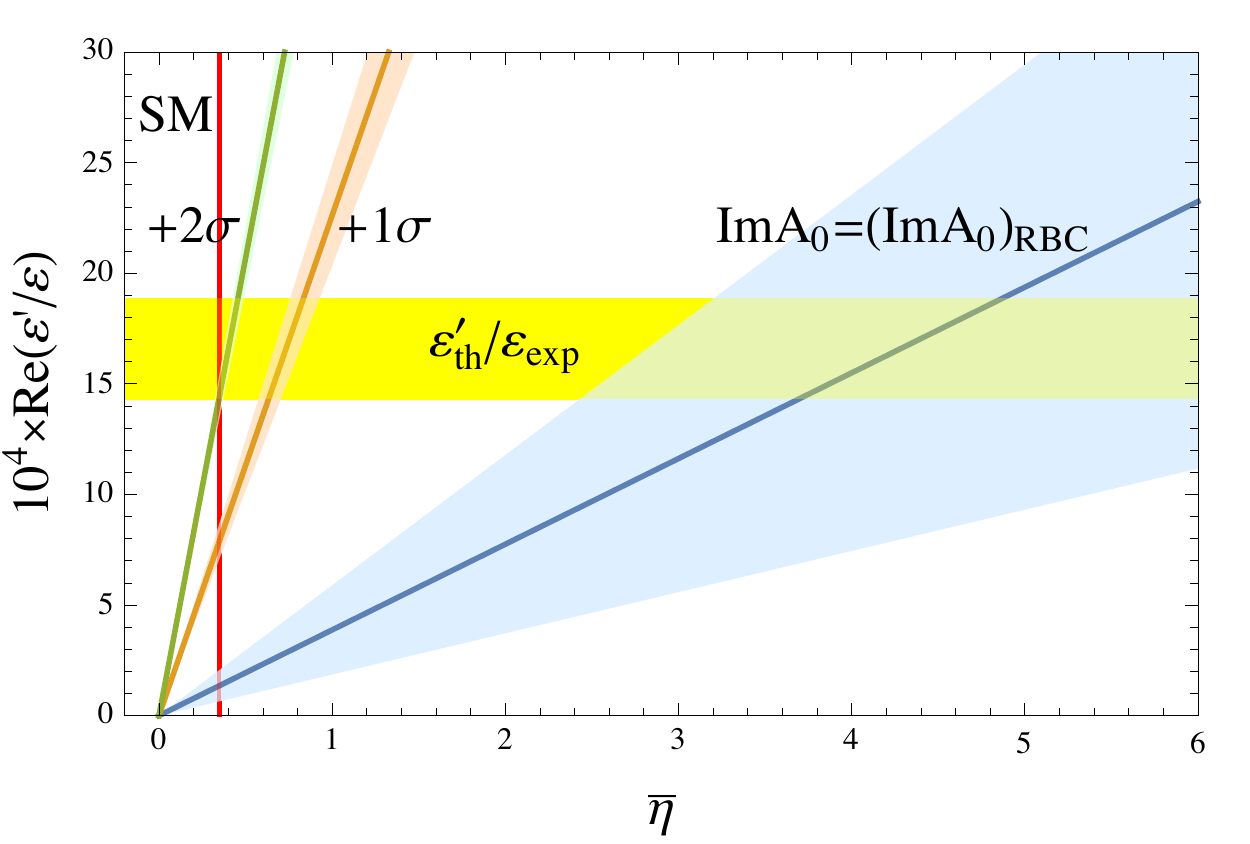}
\end{center}
\caption{Dependence of ${\rm Re}(\varepsilon^\prime/\varepsilon)$ on $\bar \eta$. The horizontal yellow band is the $1\sigma$ experimental measurement. The thin vertical red band is the $1\sigma$ determination of $\bar \eta$ from the standard unitarity triangle fit. The three lines correspond to taking ${\rm Im} A_0 = [{\rm Im} A_0]_{\rm RBC} + (0,1,2)\; \delta[{\rm Im} A_0]_{\rm RBC}$. The shaded areas around these three lines are due to all the remaining sources of uncertainties (${\rm Im} A_2$, $\delta_2$, $\delta_0$, $\phi_\varepsilon$).
\label{fig:epoe}}
\end{figure}
\begin{figure}[t]
\begin{center}
\includegraphics[width=0.99 \linewidth]{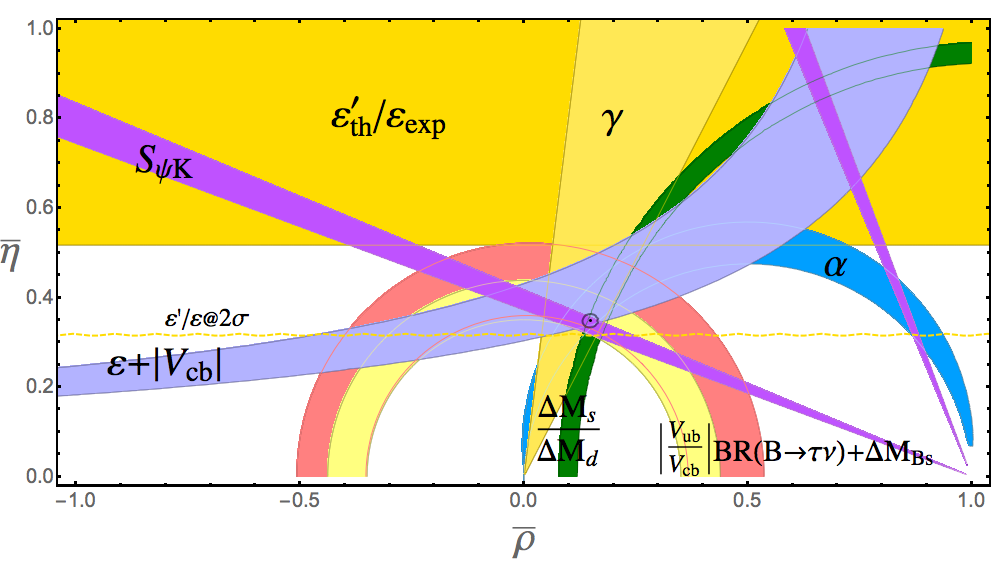}
\includegraphics[width=0.99 \linewidth]{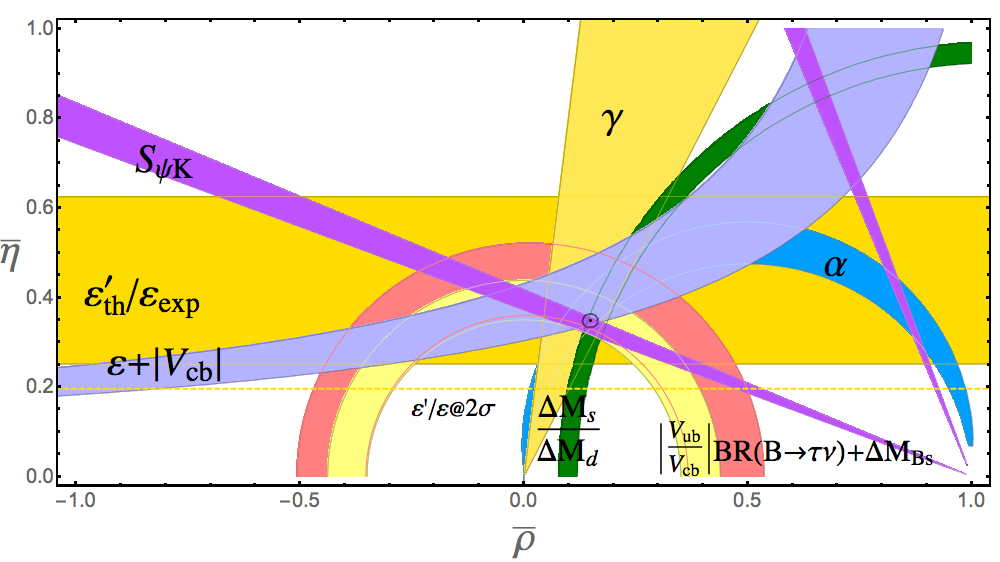} 
\end{center}
\caption{Standard unitarity triangle (SUT) fit. In the top panel we show the impact of the $\varepsilon^\prime/\varepsilon$ measurement using the most recent calculations of the $K\to (\pi\pi)_{I=0,2}$ matrix elements. In the lower panel we consider a future scenario in which the uncertainty on ${\rm Im} A_0$ and ${\rm Im} A_2$ is reduced to 18\% and 5\%, respectively (see text for more details). The ${\rm Im} A_0$ central value shifts to the value expected from the experimental determination of $\varepsilon^\prime/\varepsilon$ and the standard unitarity triangle fit. All constraints are plotted at two two sigma level with the exception of $\varepsilon^\prime_{\rm exp}/\varepsilon_{\rm exp}$ for which we show both the one and two sigma contours.\label{fig:sut}}
\end{figure}

The present experimental~\cite{Fanti:1999nm, Lai:2001ki, Batley:2002gn, AlaviHarati:1999xp, AlaviHarati:2002ye, Barr:1993rx, Gibbons:1993zq} and theoretical~\cite{Bai:2015nea} results read:
\begin{align}
{\rm Re} \left( \frac{\varepsilon^\prime_{\rm exp}}{\varepsilon_{\rm exp}} \right) & \; =
(16.6 \pm 2.3 ) \times 10^{-4} \; , \\
{\rm Re} \left( \frac{\varepsilon^\prime_{\rm th}}{\varepsilon_{\rm exp}} \right)& \; =(1.36 \pm 5.15_{\rm stat} \pm 4.59_{\rm syst} ) \times 10^{-4} 
\; .
\label{smpred}
\end{align}
Where the notation clarifies that we normalize the theoretical prediction for $\varepsilon^\prime$ to the experimental measurement of $\varepsilon$ (that has a much smaller error than the corresponding theory determination). For completeness we mention that the inclusion of the isospin breaking corrections given in Eqs.~(\ref{iso1})-(\ref{ib}) shifts the SM prediction for ${\rm Re} \left( \varepsilon^\prime/\varepsilon \right)$ to $(0.5 \pm 5.9) \times 10^{-4}$. The reduction in the total error is due to the factor $a(1-\Omega_{\rm eff}) \sim 0.87$ that multiplies the QCD penguin contribution to ${\rm Im} A_0$, resulting in a difference of $2.5 \; \sigma$ from the measured value.
\begin{figure*}[t]
\begin{center}
(a) \hskip 7.9cm (b) \\
\includegraphics[width=0.49 \linewidth]{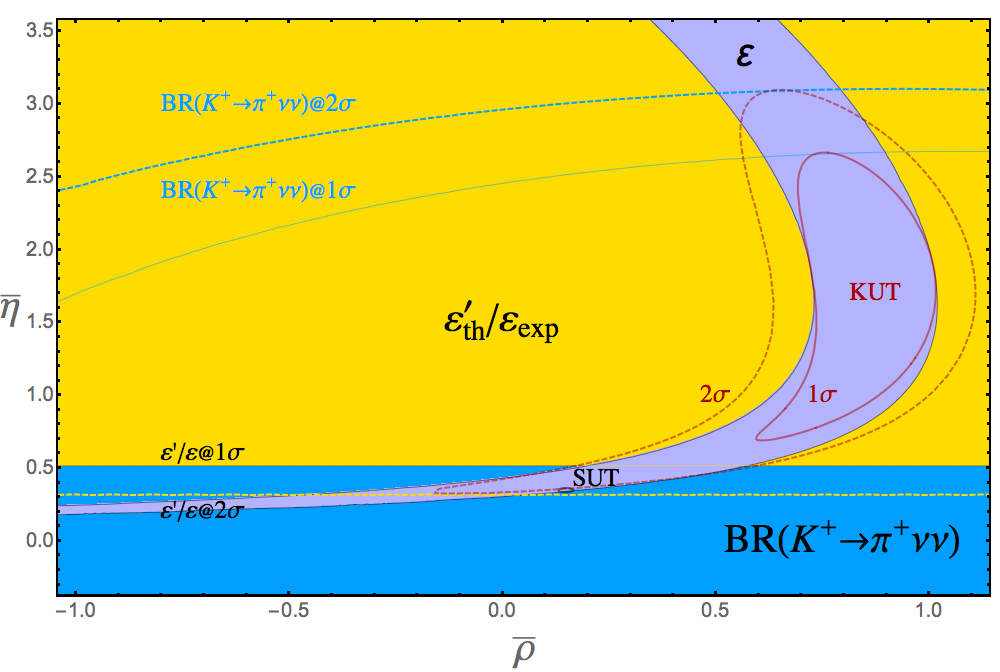}
\includegraphics[width=0.49 \linewidth]{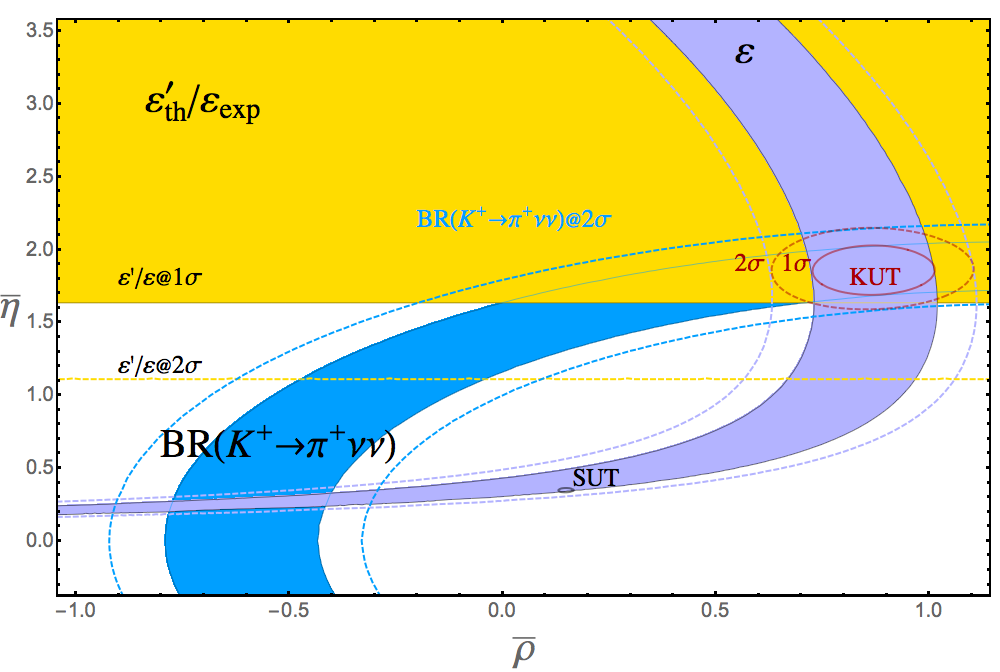} \\
(c) \hskip 7.9cm (d) \\
\includegraphics[width=0.49 \linewidth]{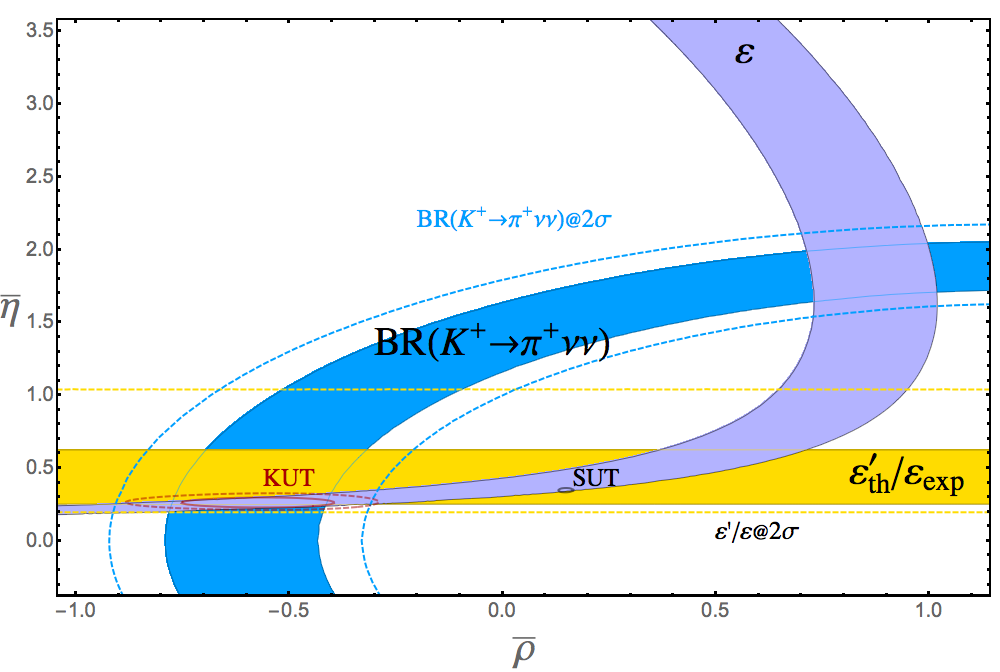}
\includegraphics[width=0.49 \linewidth]{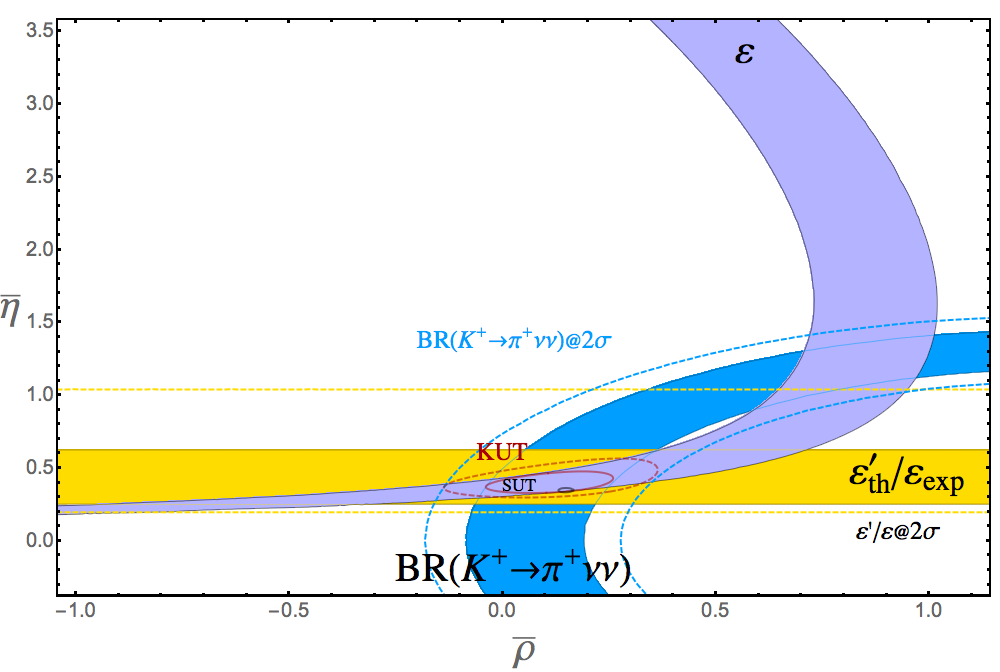}
\end{center}
\caption{Kaon unitarity triangle fits. Red contours (labelled KUT) are obtained including ${\rm BR}(K^+\to\pi^+\nu\bar\nu)$, $\varepsilon^\prime_{\rm th}/\varepsilon_{\rm exp}$, $\varepsilon$ and $|V_{cb}|$. The small black contour is the current standard unitarity triangle (SUT) fit from $B/K$ physics. In panel (a) we present the current status. The yellow area is allowed by $\varepsilon^\prime_{\rm th}/\varepsilon_{\rm exp}$ and the region below the blue curves is allowed by ${\rm BR}(K^+\to\pi^+\nu\bar\nu)$. In panels (b--d) we show the impact of future improvements on the experimental determination of ${\rm BR}(K^+\to\pi^+\nu\bar\nu)$ and on the theoretical calculation of the quantities ${\rm Im} A_0$ and ${\rm Im} A_2$ (see text for more details). In panel (b) we assume that future central values for these quantities remain unchanged. In panel (c) we consider a scenario in which ${\rm Im} A_0$ shifts to the value expected from the experimental determination of $\varepsilon^\prime/\varepsilon$ and the standard unitarity triangle fit. In panel (d) we assume, in addition, that the future experimental determination of ${\rm BR}(K^+\to\pi^+\nu\bar\nu)$ will shift to the central value of the SM prediction.
\label{fig:kfit}}
\end{figure*}

In passing we briefly note that the lattice QCD prediction in Eq.~(\ref{smpred}) differs somewhat from the result presented in Ref.~\cite{Buras:2015yba}, $(1.9 \pm 4.5) \times 10^{-4}$, which has a smaller total error. Consequently Ref.~\cite{Buras:2015yba} gets a larger deviation of $2.9\; \sigma$ from the measured value of ${\rm Re} ( \varepsilon^\prime_{\rm th}/\varepsilon_{\rm exp} )$ than the $2.1\; \sigma$ that RBC-UKQCD~\cite{Bai:2015nea} gets. This difference can be explained by a combination of (1) isospin breaking corrections and (2) the use of the operator relation, $Q_4 = Q_3 + Q_2 - Q_1$, along with the experimental information on ${\rm Re}(A_0)$ and of an estimate of the ratio of the $I=0$ matrix elements of $Q_1$ and $Q_2$ (based on the RBC-UKQCD lattice QCD results and a large-$N$ calculation) to reduce the uncertainty on $\langle Q_4\rangle_0$. Additionally, the authors of Ref.~\cite{Buras:2015yba} used the RBC/UKQCD lattice result for the dominant $\langle Q_6\rangle_0$ contribution to ${\rm Im}(A_0)/{\rm Re}(A_0)$, but expressed contributions proportional to $\langle Q_{3,5,7-10}\rangle_0$ in terms of I=0 matrix element ratios (by writing ${\rm Re}(A_0)$ in terms of $\langle Q_{1,2}\rangle_0$) that are then set to reference values inspired by large-$N$.

In both the $I=0$ and $I=2$ lattice computations, the systematic uncertainty is currently dominated by the perturbative truncation error in the computation of Wilson coefficients and the matching from the RI to MSbar scheme.  This error is of the order $\alpha_s^2(\mu)$, where $\mu$ is the scale at which one matches to perturbation theory.   By running non-perturbatively through the charm and eventually also bottom thresholds, these truncation errors will be reduced significantly in the near future.  For $\mu = 50 \; {\rm GeV}$, {\it e.g.}, perturbative truncation errors of O(1-2\%) are feasible.  Initial efforts along those lines are already in progress \cite{Frison:2014esa}.

The ten $I=0$ matrix elements are given in Table~SII of Ref.~\cite{Bai:2015nea}.  The systematic uncertainties on the individual matrix elements are obtained from those presented in Table~II of Ref.~\cite{Bai:2015nea} without including the ``Wilson coefficients'' and ``parametric errors''contributions and are about 22.5\%.

The $I=2$ matrix elements in the continuum limit and in the $(\gamma,\gamma)$ and $(q \hskip -0.15cm \slash,q \hskip -0.15cm \slash)$ RI-SMOM schemes at $3\; {\rm GeV}$ are given in the last two rows of Table~XIV of Ref.~\cite{Blum:2015ywa}. The conversion between the basis used in Ref.~\cite{Blum:2015ywa} and the standard operator basis we use is given in Eqs.~(69-70) of Ref.~\cite{Blum:2015ywa}. The conversion of these matrix elements to the $\overline{\rm MS}$ scheme at $3\; {\rm GeV}$ is achieved via the conversion matrix $C_{ij} = \delta_{ij} + \frac{\alpha_s (3 \; {\rm GeV})}{4\pi} \Delta r_{ij}$ with $i,j = 1, 7,8$. The relevant $\Delta r_{ij}$ entries are given in Table~IX and XI of Ref.~\cite{Lehner:2011fz} for the $(q \hskip -0.15cm \slash,q \hskip -0.15cm \slash)$ and $(\gamma,\gamma)$ schemes, respectively. Uncertainties associated with RI-SMOM to $\overline{MS}$ conversion are estimated by comparing the results in the $(q \hskip -0.15cm \slash,q \hskip -0.15cm \slash)$ and $(\gamma,\gamma)$ schemes, respectively.

The matrix elements that we obtain are summarized in Table~\ref{tab:bi} (for $I=0$ and $I=2$ we take $\mu=1.531 \; {\rm GeV}$ and $\mu = 3 \; {\rm GeV}$, respectively), where, as discussed above, the systematic uncertainties on the $I=0$ matrix elements are about 23\% which have been added to the respective statistical uncertainties in quadrature. The uncertainties on the $I=2$ matrix elements have been obtained by combining in quadrature the errors quoted in Table~XIV of Ref.~\cite{Blum:2015ywa} with the scale uncertainty (the latter has been defined as the difference between the $\overline{MS}$ matrix elements obtained via an intermediate $(q \hskip -0.15cm \slash,q \hskip -0.15cm \slash)$ or $(\gamma,\gamma)$ RI-SMOM scheme).

The remaining statistical and systematic uncertainties in the isospin symmetric limit are amenable to improvement by a straightforward numerical effort.  We therefore believe that an error on individual $I=0$ operator matrix elements of order 5\%-10\% is achievable within five years given sufficient computational resources.  Using current central values of matrix elements, the corresponding propagated error on ${\rm Im}(A_0)$ is O(10-20\%). 
\begin{figure}[t]
\begin{center}
\includegraphics[width=0.99 \linewidth]{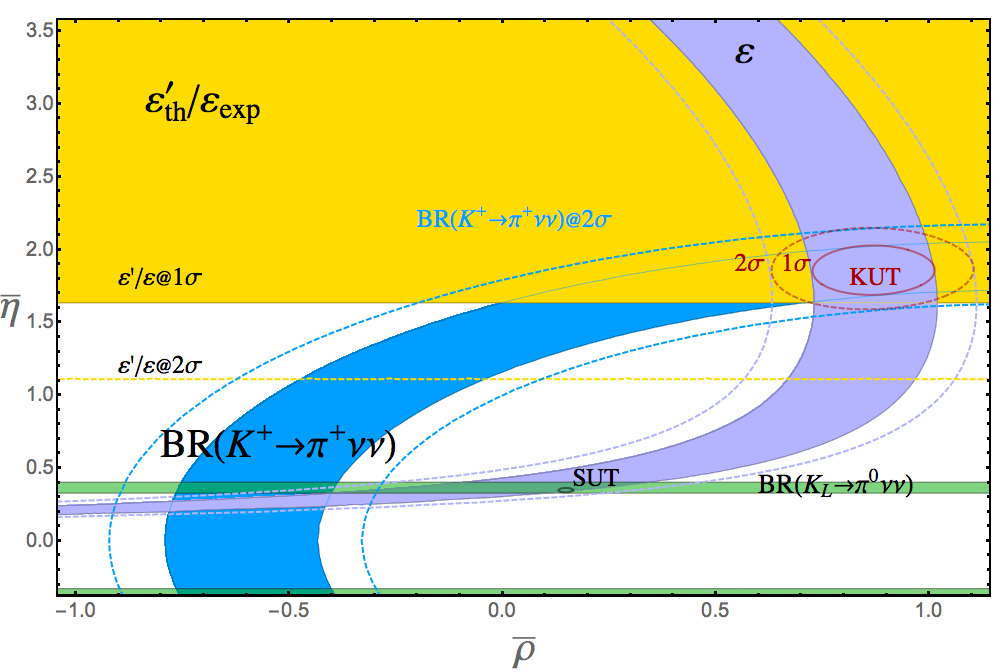}
\end{center}
\caption{Impact of a future measurement of ${\rm BR} (K_L\to \pi^0\nu\bar\nu)$ assuming SM central values with a $10\%$ uncertainty~\cite{Komatsubara:2012pn, Shiomi:2014sfa}.
\label{fig:klpnn}}
\end{figure}

Motivated by this discussion, we assume that the uncertainties on the $I=0,2$ matrix elements will reduce to $\delta {\rm Im}(A_2) = 5\%$ and $\delta {\rm Im}(A_0) = 10\%_{\rm stat} + 15\%_{\rm syst} = 18\%$ on a time-scale of five years. Note that these error estimates are given with respect to the central values in Eqs.~(\ref{rbca2}) and (\ref{rbca0}). Allowing for fluctuations of the central values of the matrix elements that control ${\rm Im} (A_0)$ the expected future theoretical errors on this quantity is $\delta {\rm Im}(A_0) \simeq 0.34 \times 10^{-11} \; {\rm GeV}$.

Before concluding this section let us comment on the treatment of the correlations between the errors on the matrix elements listed in Table~\ref{tab:bi}. In the extraction of ${\rm Im} A_I\; (I=0,2)$ given in Eqs.~(\ref{rbca2}) and (\ref{rbca0}) all statistical correlations have been taken into account; systematic uncertainties for ${\rm Im} A_0$ were ascribed to individual operator matrix elements and treated as uncorrelated (correlations tended to reduce the errors). On the other hand, we neglect correlations between the ${\rm Im} A_2$ and ${\rm Im} A_0$ errors. Given the difference in the uncertainties on these two matrix elements, correlations can be safely neglected at present but will need to be included when the total error on ${\rm Im} A_0$ will become small enough.

\section{Indirect CP-violation and \texorpdfstring{$\varepsilon$}{}}
\label{sec:ek}
The basic expression for $\varepsilon$ is
\begin{align}
 \varepsilon  = & \;  e^{i\phi_\varepsilon} \frac{G_F^2 m_W^2 f_K^2 m_K}{12 \sqrt{2} \pi^2 \Delta m_K^{\rm exp}} 
 \hat B_K  \kappa_\varepsilon \; {\rm Im} \Big[ \nonumber \\
& \eta_1 S_0 (x_c) \left( V_{cs}^{} V_{cd}^* \right)^2 + \eta_2 S_0 (x_t) \left( V_{ts}^{} V_{td}^* \right)^2  \nonumber \\
&  + 2 \eta_3 S_0 (x_c,x_t) V_{cs}^{} V_{cd}^*  V_{ts}^{} V_{td}^*\Big] \; , \label{ek}
\end{align}
where the numerical inputs we use are summarized in Table~\ref{tab:ekinputs}. The quantity $\kappa_\varepsilon$ summarizes the impact of long distance effects and can be extracted from the knowledge of ${\rm Im} \; A_0$ and from an estimate of the long distance contributions to $\Delta m_K$. Following Ref.~\cite{Buras:2010pza}, we have:
\begin{align}
\kappa_\varepsilon & \; = \sqrt{2} \sin(\phi_\varepsilon) \left( 1 + \frac{\rho}{\sqrt{2} \left|\varepsilon_{\rm exp}\right|} \; \frac{{\rm Im} (A_0)}{{\rm Re} ( A_0)} \right)
\end{align}
where $\rho = 0.6 \pm 0.3$.  Using the most recent RBC determination of ${\rm Im}(A_0)$ and $\phi_\varepsilon$ of Eq.~\eqref{eqn:phieps}, we obtain $\kappa_\varepsilon = 0.963 \pm 0.014$ (See also the analysis presented in Ref.~\cite{Ligeti:2016qpi}).

The single largest remaining parametric uncertainty on $\varepsilon$ is due to $|V_{cb}|$. The latter determines the parameter $A$ of the CKM Wolfenstein parametrization that enters $\varepsilon$ to the fourth power (via $V_{ts}^{} V_{td}^* \propto A^2$). Presently both inclusive and exclusive $b\to c \ell\nu$ decays lead to a $\sim 2\%$ error on $|V_{cb}|$; unfortunately a $2.5\sigma$ tension between these two determinations lead, via standard PDG rescaling, to a $2.5\%$ error on the averaged result. Unless this tension signals some form of new physics, the most probable avenue to improve the total uncertainty on $\varepsilon$ is hoping that future experimental and theoretical results for $b\to c\ell\nu$ transitions will resolve this tension thus leading to a final sub-percent error on $|V_{cb}|$. For completeness we mention that an alternative approach to unitarity triangle studies that does not make use of semileptonic decays has been proposed in Ref.~\cite{Lunghi:2009ke}.

\begin{figure*}[t]
\begin{center}
(a) \hskip 7.9cm (b) \\
\includegraphics[width=0.49 \linewidth]{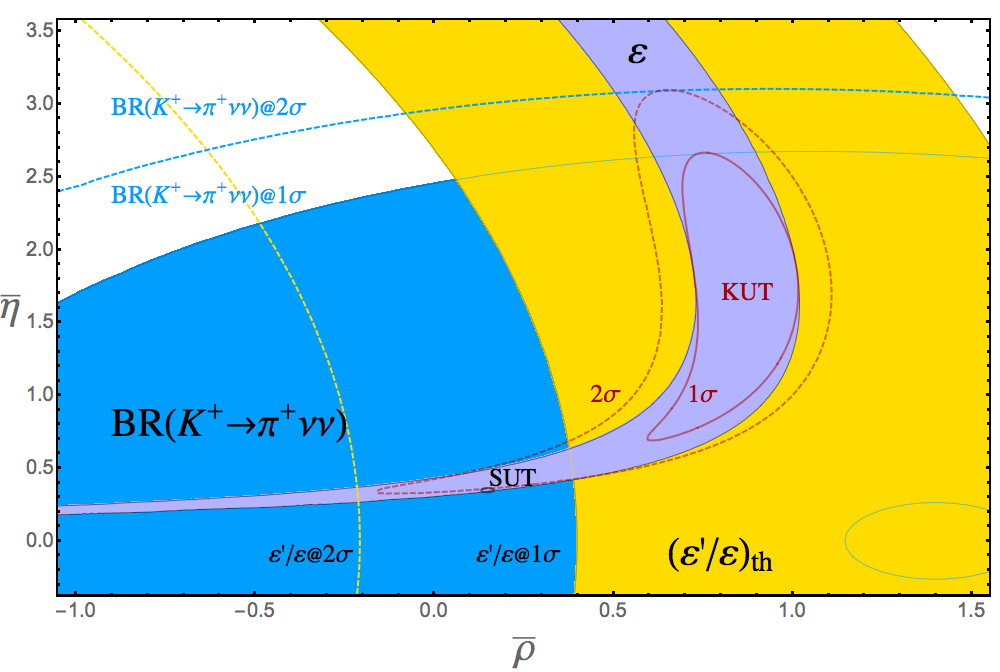}
\includegraphics[width=0.49 \linewidth]{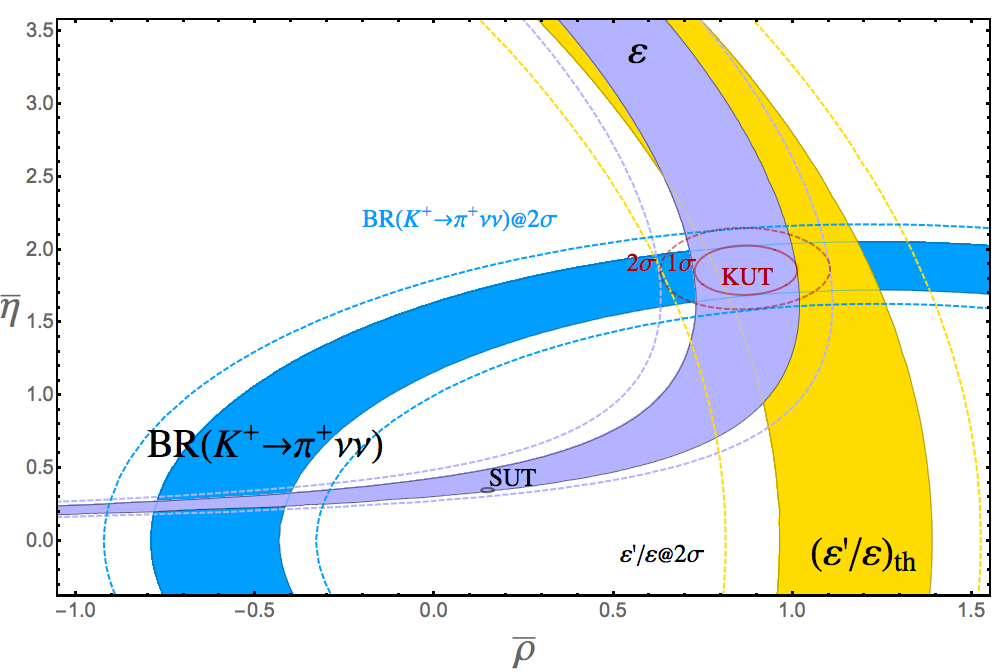} \\
(c) \hskip 7.9cm (d) \\
\includegraphics[width=0.49 \linewidth]{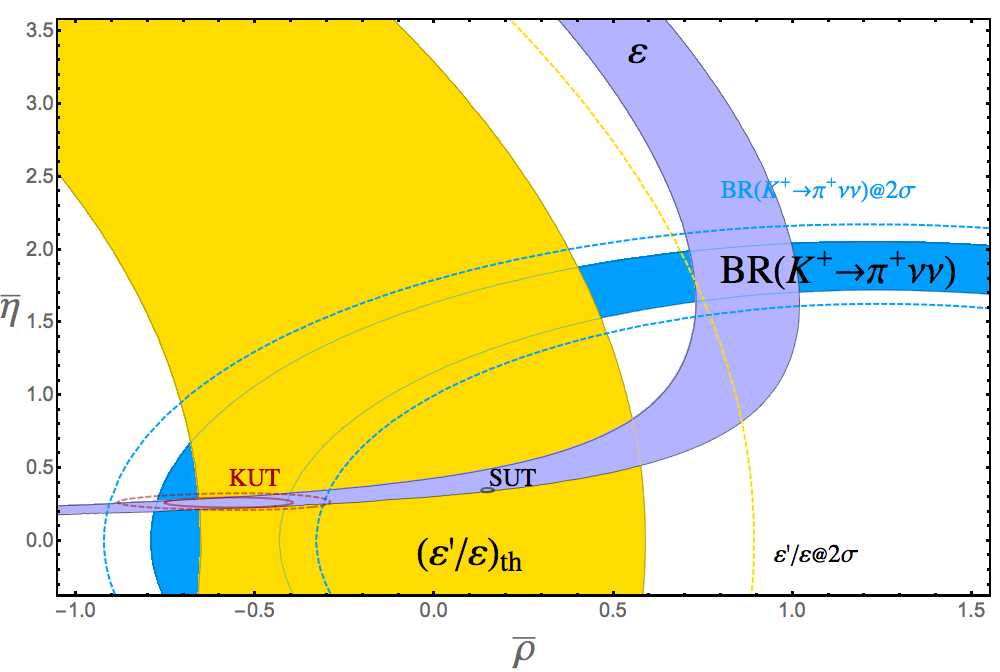}
\includegraphics[width=0.49 \linewidth]{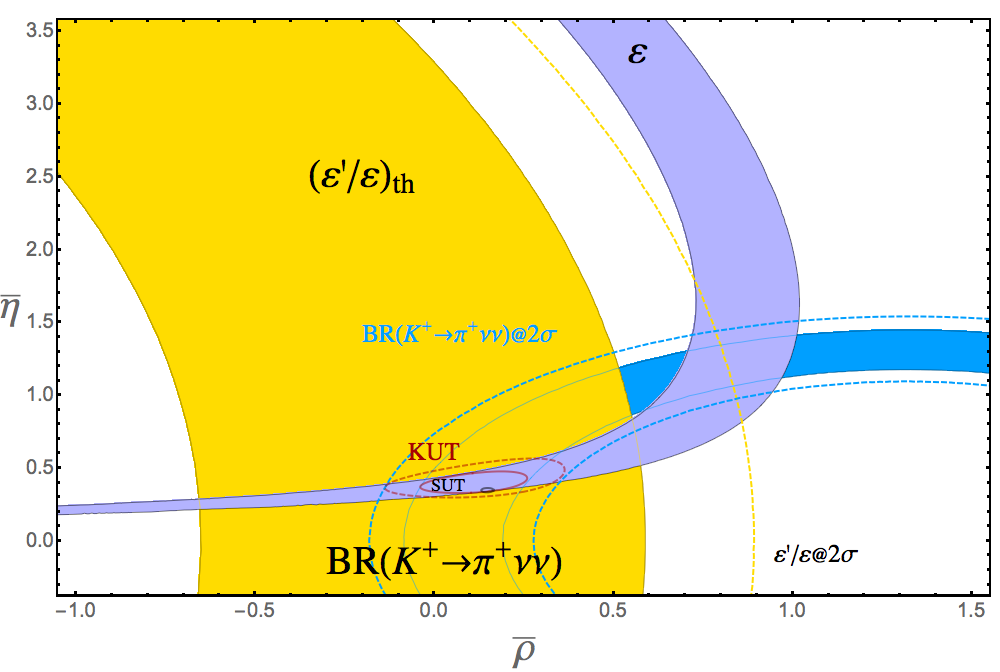}
\end{center}
\caption{Kaon unitarity triangle fits using $\varepsilon^\prime_{\rm th}/\varepsilon_{\rm th}$ rather than $\varepsilon^\prime_{\rm th}/\varepsilon_{\rm exp}$. See the caption in Fig.~\ref{fig:kfit} for further details. \label{fig:kfit-th}}
\end{figure*}

\section{Results}
\label{sec:results}
In Fig.~\ref{fig:epoe} we show the dependence of ${\rm Re} (\varepsilon^\prime/\varepsilon)$ on $\bar \eta$ for different choices of the ${\rm Im}(A_0)$ central value. Note how the uncertainty on this matrix element completely dominates the total uncertainty. 

In Fig.~\ref{fig:sut} we present the standard unitarity triangle (SUT) fit obtained using $B$ and $K$ physics measurements. All the inputs used in the fit are taken from Refs.~\cite{Agashe:2014kda, Aoki:2013ldr, Amhis:2014hma} In particular, we adopt $|V_{cb}| = 40.41(94) \times 10^{-2}$ and $|V_{ub}| = 3.68(26) \times 10^{-3}$ obtained from a combination of inclusive and exclusive determinations of these CKM elements (because of the slight tensions between inclusive and exclusive $|V_{qb}| \; (q=u,c)$, the errors on the weighted averages are rescaled following the standard procedure described in Ref.~\cite{Agashe:2014kda}). In the upper panel we show the present constraints imposed by $\varepsilon^\prime_{\rm th}/\varepsilon_{\rm exp}$ (this notation underscores that in the theoretical prediction for the $\varepsilon^\prime/\varepsilon$ ratio the denominator is taken from experiment and not calculated using the inputs listed in Table~\ref{tab:ekinputs}). In the lower panel we entertain a future scenario in which the uncertainty on ${\rm Im} A_0$ is 18\%, as discussed above.  Its central value is assumed to shift to what is necessary to reproduce the experimental determination of $\varepsilon^\prime/\varepsilon$ (note that even though $\varepsilon^\prime/\varepsilon_{\rm exp}$ is proportional to $\bar \eta$, the $\chi^2$ minimized over every other parameter is not a symmetric function of $\bar \eta$). If the future ${\rm Im} A_0$ central value does not shift, the $\varepsilon^\prime/\varepsilon$ allowed region is $\bar \eta \gtrsim 1.6$ (see Fig.~\ref{fig:kfit}b) implying a very strong tension with the standard fit.

In Fig.~\ref{fig:kfit} we present the Kaon unitarity triangle fits (KUT) in various scenarios. In this fit we use only inputs from Kaon physics with the exception of tree--level determinations of $|V_{c b}|$ from inclusive and exclusive $b\to c\ell\nu$ decays. The red contours are obtained including ${\rm BR}(K^+\to\pi^+\nu\bar\nu)$, $\varepsilon^\prime/\varepsilon$, $\varepsilon$, and $|V_{cb}|$ (from an average of inclusive and exclusive decays). The small black contour is the current standard unitarity triangle fit from $B/K$ physics. 

In Fig.~\ref{fig:kfit}a we present the current status of this fit. The yellow area is allowed by $\varepsilon^\prime/\varepsilon$ and the region below the blue curves is allowed by ${\rm BR}(K^+\to\pi^+\nu\bar\nu)$. 

In Figs.~\ref{fig:kfit}b--\ref{fig:kfit}d we show the impact of future improvements on the experimental determination of ${\rm BR}(K^+\to\pi^+\nu\bar\nu)$ and on the theoretical calculation of the quantities ${\rm Im} A_0$ and ${\rm Im} A_2$. In particular, in panel (b) we assume that future central values for these quantities remain unchanged and that experimental and theoretical uncertainties on ${\rm BR}(K^+\to\pi^+\nu\bar\nu)$ and ${\rm Im} (A_{0,2})$ reduce as discussed above. In panel (c) we consider a scenario in which ${\rm Im} A_0$ shifts to the value expected from $\varepsilon^\prime_{\rm exp}/\varepsilon_{\rm exp}$ and the standard unitarity triangle fit. In panel (d) we assume, in addition, that the future experimental determination of ${\rm BR}(K^+\to\pi^+\nu\bar\nu)$ will shift to the central value of the SM prediction.

Finally in Fig.~\ref{fig:klpnn} we show the impact of a future measurement of ${\rm BR} (K_L \to \pi^0\nu\bar\nu)$ at the SM level with 10\% uncertainty.

\begin{figure}[t]
\begin{center}
\includegraphics[width=0.99 \linewidth]{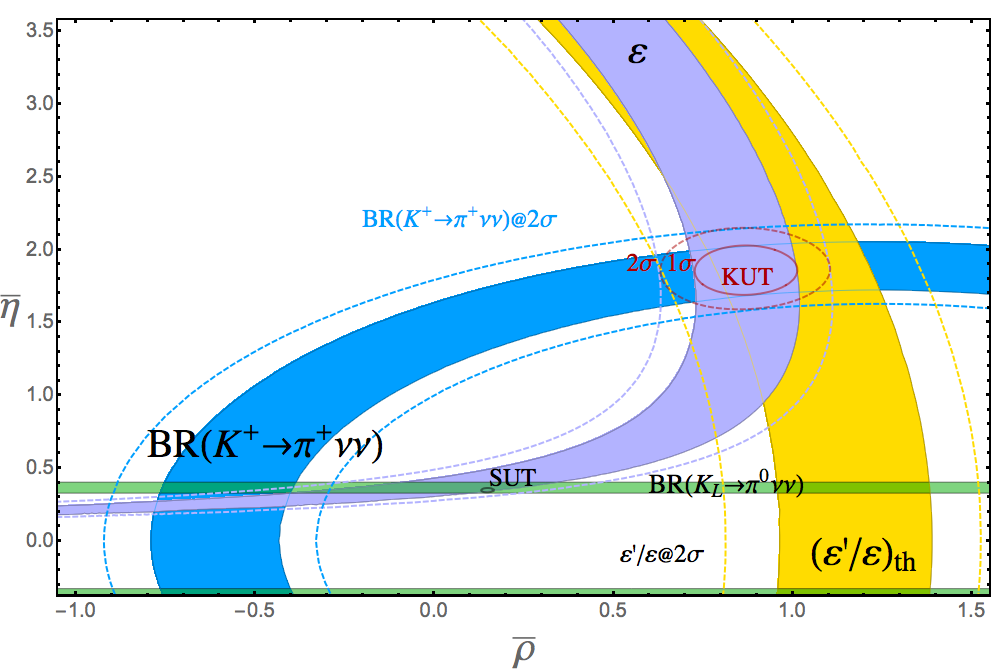}
\end{center}
\caption{Impact of a future measurement of ${\rm BR} (K_L\to \pi^0\nu\bar\nu)$ assuming SM central values with a $10\%$ uncertainty~\cite{Komatsubara:2012pn, Shiomi:2014sfa}.
\label{fig:klpnn-th}}
\end{figure}

In Figs.~\ref{fig:kfit-th} and \ref{fig:klpnn-th} we present the very same fits but display $\varepsilon^\prime_{\rm th}/\varepsilon_{\rm th}$ rather than $\varepsilon^\prime_{\rm th}/\varepsilon_{\rm exp}$. The reason for considering the theoretical ratio $\varepsilon^\prime_{\rm th}/\varepsilon_{\rm th}$ is that it is independent of the unique SM CP violating parameter $\bar\eta$ (both numerator and denominator being proportional to $\bar\eta$). As precise theoretical calculations of this ratio become available, direct comparison with the corresponding experimental ratio can provide a new CP-conserving observable that can be used to constrain the relevant combination of CKM parameters in the UT-fit.

\section{Conclusions and Outlook}
\label{sec:conclusions}
In this work we have tried to draw attention to the significant progress that lattice QCD methods have recently made for quantitatively addressing non-perturbative effects in several $K$-decays such as $K \to \pi \pi$ and the direct CP-violation parameter ${\rm Re} \left(\varepsilon^\prime/\varepsilon\right)$, the long-distance contribution to $\varepsilon$, and rare $K$-decays; of particular interest will be future lattice QCD determination of the long distance contributions to the quantity $P_c(X)$ that contains charm quark contributions to ${\rm BR} (K^+\to\pi^+\nu\bar\nu)$~\cite{Isidori:2005tv}.

This means in the near future we will be able make better use of experimental data, existing and forthcoming, to better constrain the SM and search for new effects. In particular, it appears that we can start to construct a unitarity triangle based primarily on $K$-physics.  With improvements in the lattice calculations that are on the horizon and with results from forthcoming $K$-experiments a tighter KUT should soon become available. It would be very valuable to compare the solution of such an improved KUT with the Standard Unitarity Triangle (SUT) coming primarily from $B$-physics.

In particular it now seems realistic that lattice calculations can reduce the errors on ${\rm Re}\left({\varepsilon^\prime/\varepsilon}\right)$ to less than around 11\% of the current experimental central value in about 5-years time. It may therefore be timely for the experimental community to plan an improved determination of  ${\rm Re}\left({\varepsilon^\prime/\varepsilon}\right)$, the current experimental errors on that quantity being around 15\%.

\section{Acknowledgment}
C.L. and A.S. want to thank their collaborators from RBC-UKQCD for discussions and ecouragement. The work of CL and A.S. is supported in part by US DOE Contract \#AC-02-98CH10886(BNL). The work of E.L. is supported in part by the United States Department of Energy under grant number DE-SC0010120.

\bibliography{kfit.bib}

\begin{thebibliography}{80}
\bibitem [{LL()}]{LL}%
  \BibitemOpen
  \href@noop {} {}\bibinfo {note} {{We want to thank Laurence Littenberg for
  discussions regarding this background}}\BibitemShut {NoStop}%
\end{thebibliography}

\end{document}